# Transformation of hydrogen bond network during the $CO_2$ clathrate hydrate dissociation

Kirill Gets[1,2,*], Vladimir Belosludov[1,2], Ravil Zhdanov[1,2], Yulia Bozhko[1,2], Rodion Belosludov[3], Oleg Subbotin[1,2], Nikita Marasanov[1], Yoshiyuki Kawazoe[4,5]

[1]Novosibirsk State University, 1, Pirogova str., Novosibirsk, 630090, Russia
[2]Nikolaev Institute of Inorganic Chemistry SB RAS, 3, Acad. Lavrentiev Ave., Novosibirsk, 630090, Russia
[3]Institute for Materials Research, Tohoku University, 2-1-1 Katahira, Aoba-ku, Sendai, 980–8577, Japan
[4]New Industry Creation Hatchery Center, Tohoku University, 4-4-6 Aramaki aza Aoba, Aoba-ku, Sendai, 980-8579, Japan
[5]SRM institute of Science and Technology, SRM Nagar, Kattankulathur, Kancheepuram District, Tamil Nadu, 603 203, India

**Abstract**

The object of this study is the kinetic process of solid-liquid first-order phase transition – melting of carbon dioxide CS-I hydrate with various cavity occupation ratios. The work was done within a framework of study on the local structure of water molecules. These include the time depending change of the short-range order at temperatures close to the melting point and comparison with hexagonal ice structure. Using molecular dynamics method, dependencies of the internal energy of the studied systems on the time of heating were found. Jumps in the internal energy of solids in the range at 275-300 K indicate a phase transition. The study of oxygen-oxygen radial distribution and hydrogen-oxygen-oxygen mutual orientation angles between molecules detached at no more than 3.2 Å allowed to find the H-bond coordination number of all molecules and full number of H-bonds and showed the instant (less than 1 ns) reorganization of short-range order of all molecules. The structure analysis of every neighbor water molecules pairs showed the ~10-15% decrease of H-bond number after the melting whereas all molecules form single long-range hydrogen bond network. The analysis of hydrogen bond network showed the minor changes in the H-bond interaction energy at solid-liquid phase transition.

**Keywords**

Gas hydrate, phase transition, molecular dynamics, hydrogen bond network, short-range order.

# 1. Introduction

Gas hydrates are crystalline water-based compounds. Their structure consists of water as the host molecules and trapped inside gas as guest molecules. The most common structures are: CS-I, CS-II and HS-III. The cavities in host lattice consist of hydrogen bonded tetragons, pentagons and hexagons. For example, the CS-I hydrate consists of large cavities (twelve pentagons and two hexagons, $5^{12}6^2$) and small cavities (twelve pentagons, $5^{12}$). Each cavity can hold a number of various guest molecules depends on their size [1]. Thermodynamic stability of clathrate hydrates depends on *p-T* conditions as well as type of guest molecules inside of hydrate cavities. [2, 3].

Gas hydrates are considered as attractive media to store and sequester carbon dioxide ($CO_2$) in the solid form [4-6]. This problem is considered as part of a task of replacing methane in methane hydrate deposits, which are associated with a high threat of methane in the greenhouse effect in the case of methane hydrate melting and gas release [7-9].

Development of technologies for carbon dioxide utilization and storage in hydrate form requires a clear understanding of the processes of formation and dissociation of $CO_2$ hydrates. Recently, the molecular dynamic (MD) studies of gas hydrate described the dissociation processes by gradual destruction of hydrate structure at hydrate-liquid water interface (dissolution) [10-12]. The other possibility of the melting caused by hydrate heating only gives an idea of the structural changes in the host lattice close to the solid-liquid phase transition. This is an attempt to find the answer to the fundamental question: whether the melting of solid is a first-order phase transition or a continuous process [13].

Melting of a hydrate can be considered as a heterogeneous/homogeneous process in a presence/absence of imperfections [14], such as, for example, empty cavities. The results of MD simulations of ideal crystals shows the 20-30% increase in value of melting temperature in comparison to experimental data [15, 16]. However it can diminish the high value of melting temperature by the increasing the system size [14, 17].

Molecular dynamics is a powerful and precise theoretical method to study both macroscopic properties and microscopic structure and to describe a relationship between them. However, the study of the kinetic processes that taken place at gas hydrate formation and dissociation are limited [10, 11]. The most common water models for describing hydrates melting are: TIP4P [18], SPC/E [19], TIP4P-Ew [20], TIP4P/Ice [21], TIP4P/2005 [22]. Any of these potentials are able to give a qualitative description of the processes of formation and decomposition of gas hydrates. The TIP4P and SPC/E potentials are preferred for gas hydrate dissociation kinetic studying [12, 23, 24], and the rest of potentials for the studying of hydrate formation [17, 25]. A more detailed description can be found in the work [26].

Melting of gas hydrates was studied in a series of works [27-32] which are devoted to the study of thermodynamic properties, such as free energy or chemical potential of hydrate systems [33, 34], effects of "help" gases or inhibitors on hydrate stability [35-37], but at the moment there is no description of the melting of hydrates within the framework of the hydrogen bond network reorganization and changes of energy and structure properties of every close $H_2O$ pair (short-range order of every molecule).

Our work is devoted to study the kinetics processes of melting $CO_2$ hydrate in the absence of liquid water and with constant uniform heating of the hydrate and hence to study the phase transition on molecular level. The local reorganization can be generalized and described in the terms of hydrogen bond network. The hydrogen bond network unites all water molecules. Its local structure [38] and collective dynamics [39] are responsible for water properties. The process of clathrate hydrate melting rises a problem of study the transformation of hydrogen bond network on the liquid water − clathrate hydrate phase transition [40, 41] as well as on the surface of the resultant gas bubbles.

In this paper, we propose a description of melting process via the structural analysis of neighboring water molecule ensemble ($R_{O-O} < 3.2$ Å) and their interaction energy. The study is conducted on a series of snapshots (instantaneous configurations) obtained by MD during the modeling of crystalline hexagonal ice ($I_h$), CS-I hydrate with 0%, 25% and 50% ratio of cavity occupation by the $CO_2$ molecules close to the melting point and. The aim is to study the changing of structural (short-range order) and energy properties of $CO_2$ hydrate and ice during the melting process under the heating.

## 2. Calculation details

To model the $CO_2$ hydrate melting and compare its structure characteristics with melting Ih ice, the MD method within the approximation of a rigid molecule was used (LAMMPS software package, [42]). The interaction between water molecules was described by the modified simple point charge extended (SPC/E) potential [19]. It was chosen because it qualitatively describes the kinetics of dissociation [12, 23, 24]. The modified version was used according to good description of thermodynamics for these phases in our previous studies [43, 44]. The short-range interaction was described by Lennard-Jones potential with parameter σ = 3.1556 Å and the energy parameter ε = 0.65063 kJ mol−1. The charges on hydrogen ($q_H$ = +0.4238|e|) and on oxygen ($q_O$ = −0.8476|e|) atoms were obtained to describe the long-range Coulomb interaction. Other parameters were set

original [19]. The $CO_2$ molecules were considered as single particle and their parameters were $\sigma = 4.0$ Å and the energy parameter $\varepsilon = 1.5798$ kJ mol$^{-1}$ [45-47].

The supercell cell of 10 x 10 x 10 unit cells of CS-I hydrate with 0%, 25% and 50% rate of single carbon dioxide occupation of cavity was chosen for the calculations. The procedure for obtaining the structure of crystalline ice $I_h$ has been described previously in detail [48].

The structure changing during the melting process the hydrate and the ice were modeled using an NPT ensemble. The pressure was set to 1 bar, the initial temperature was set to 275 K and the heating rate differs from ~0.05 K/ns for CS-I structures, and ~0.1 K/ns for $I_h$ ice. During the simulation, the coordinates of all atoms within the model were recorded at each 5 fs for empty CS-I hydrate, 100 fs for 25% and 50% cavity filled CS-I hydrate, and 200 fs for $I_h$ ice, determining the instantaneous configurations of the systems (snapshots). The temperature and pressure of the model systems are controlled by Nose-Hoover style non-Hamiltonian equations of molecular motion [49].

To study hydrogen bond network structure characteristics the analysis of a series of snapshots before the melting point and after it has been performed. This includes:
a) determining the dependence of the distribution of the number of closest O···O pairs ($R_{O-O} < 3.2$ Å) on the distance between them, on the mutual orientation angles ∠H−O···O. Defining the H-bond formation between the O···O pair using the geometric criteria ($R_{O-O} < 3.2$ Å, ∠H−O···O < 30º) [50]. This criteria were chosen due to slight differences between energy, geometric and hybrid criteria [51, 52];
b) determining the dependence of the distribution of the number of $CO_2$···$CO_2$ on the distance between them;
c) determining the dependence of the number of H-bonds per molecule as a function of time and temperature and the average number of molecules linked by H-bonds in the nearest environment (short-range order) of all molecules;
d) calculating the interaction energies between neighbor water molecules as a sum of Coulomb and van der Waals (Lennard-Jones) potentials.

## 3. Results and discussion

The time dependence of caloric curves for ice and $CO_2$ hydrate are presented in Fig. 1. The caloric curve represents the internal energy of the system i.e. the sum of potential and kinetic energies. The empty CS-I structure collapsed almost at the beginning of simulation at 275 K. The hydrate with 25% occupation melted after 15 ns at 275 K and at 50% of occupation, the hydrate melted at 300 K after the 487 ns of heating. As the solids melt the internal energy rises in absolute value by ~10%, except the melting of CS-I hydrate with 50% cavity occupation ratio, where the energy increasing value is ~14% that could be influenced by the presence of high number of carbon dioxide molecules as well as higher temperature. These results agree well with the prediction calculated from the Clapeyron equation, the experimental data on $CO_2$ clathrate hydrate dissociation [53] and the simulation results of melting of water in solid phase [54, 55].

The observed jump of internal energy is a sign of first-order phase transition: the melting process take less than one nanosecond in the comparison to the surface melting, i.e in the presence of liquid water on the surface. In that case, the dissociation could take 20–50 ns [10] for the cell of the same size and more for larger cell model, because first-order phase transition taking place through the full volume and should characterizes by the intermittent processes that are continuous at the surface melting. A short time of phase transition (structure transformation time) has collective nature and it takes place due to the fact that the network of hydrogen bonds even in the liquid phase is dynamic [39] and its structure changes have collective nature as was found [56-59].

Transition instantaneity will also be clearly seen at the following figures proving the first-order phase transition.

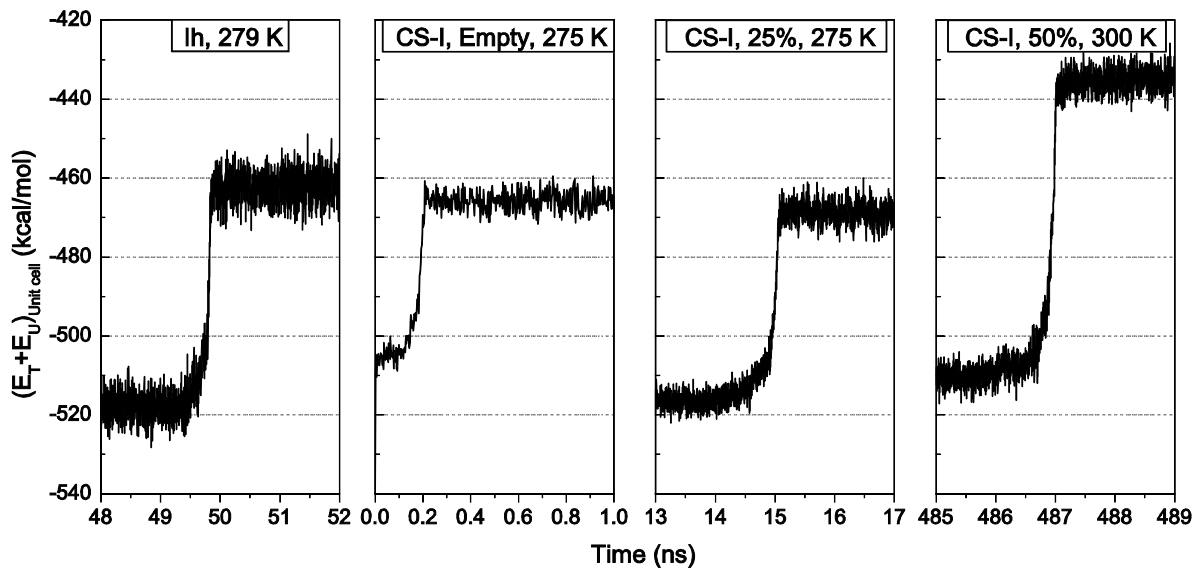

Fig.1. Caloric curves for CS-I hydrates with different occupancy rates at 275 K and 300 K and $I_h$ ice at 279 K. Dependence of kinetic and potential energy sum on time. Caloric curve of ice was normalized to the values as they consist of 46 molecules.

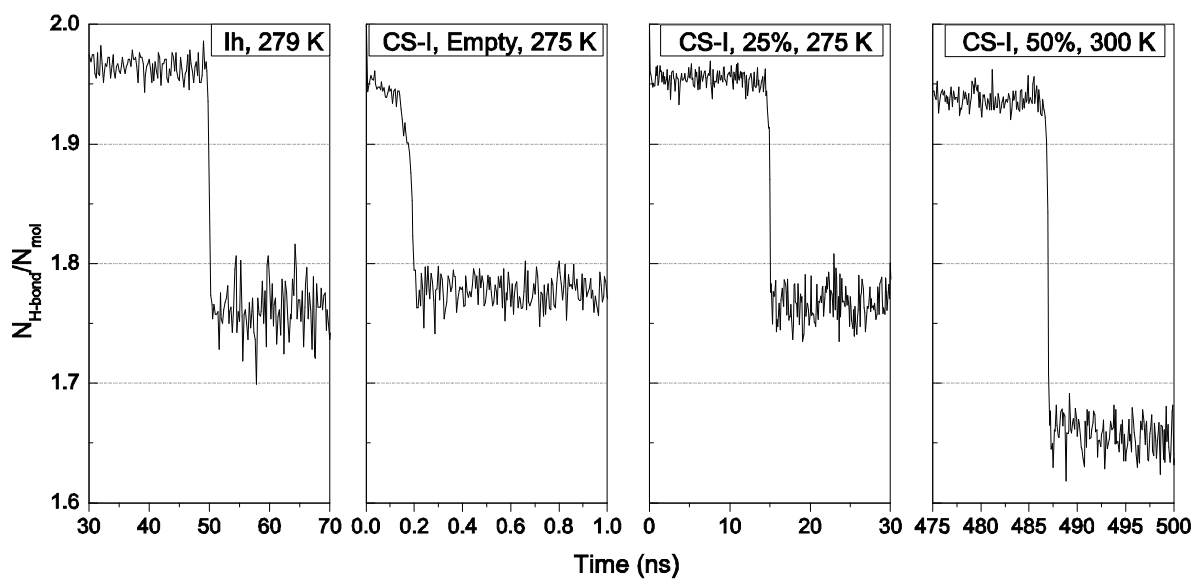

Fig. 2. Dependence of H-bond number $N_{H-bonds}$ normalized to number of molecules $N_{mol}$ as a function of time for CS-I hydrates with different occupancy rates at 275 K and 300 K and $I_h$ ice at 279 K.

The study of molecule pairs, which mutual geometry satisfies the hydrogen bond geometrical criteria, made it possible to calculate the number of hydrogen bonds in hydrate and ice before and after melting point. These pairs form the H-bond network and the dependence of their normalized number on time is presented in Fig. 2. Before the simulation started (without thermal motion) every water molecule of $I_h$ ice and hydrates has exactly 4 molecules in the short-range order. It corresponds to $N_{H-bonds}/N_{mol}$ exactly equals 2. However, due to the thermal motion some H-bonds in the solid structure are broken and $N_{H-bonds}/N_{mol}$ is lower than 2.

It can be seen that the number of H-bonds varies with time, but is near the average value. The change in the number of hydrogen bonds with time is connected with the constant rebuilding of the local order of each water molecule — the change in the angles of mutual orientation and distances between molecules due to thermal motion. As it was shown earlier, the transition from the solid to the liquid phase takes less than one nanosecond: ubiquitous rebuilding of short-range order structure clearly indicates a collective process. For melted ice and hydrate average value of $N_{H\text{-bonds}}/N_{mol}$ depends on temperature and lowers with temperature increase, however the effect of $CO_2$ molecule presence improves the stability of hydrate structure with increase of cavity occupancy rate. Another point is the melted state when $CO_2$ molecules are wedged in $H_2O$ molecule short-range order, preventing the formation of H-bonds that can be seen for melted 50% occupied hydrate. Empty and 25% occupied CS-I and $I_h$ ice structures are in good agreement with results obtained by Pauling [60] which states that ice melting leads to the breaking of ~15% of H-bonds.

We found that in every snapshot (at any time) of water system, all water molecules form a dynamic H-bond network. Thus, the H-bond network is preserved when solid water melts.

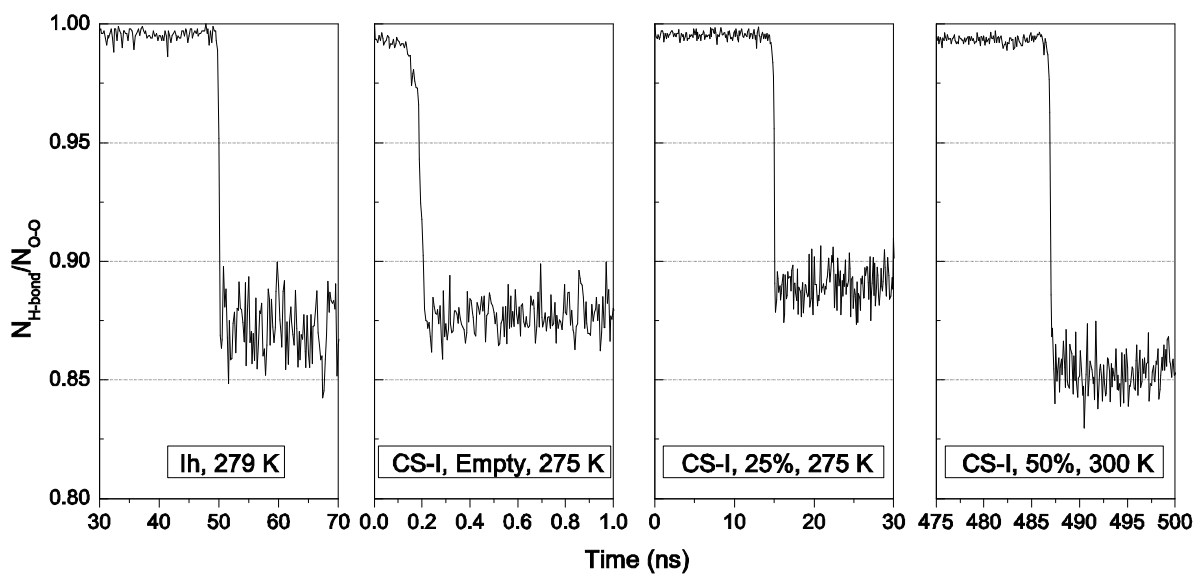

Fig. 3. Dependence of H-bond number $N_{H\text{-bonds}}$ normalized to number of O···O pairs ($R_{O-O} < 3.2$ Å) $N_{O-O}$ as a function of time for CS-I hydrates with different occupancy rates at 275 K and 300 K and $I_h$ ice at 279 K.

Geometry changes that H-bond network undergoes upon melting of $CO_2$ CS-I hydrate and $I_h$ ice can be estimated by changes in the H-bond lengths ($R_{O-O}$) and the mutual orientation angles ($\angle H-O\cdots O$) between neighbor $H_2O$ molecules ($R_{O-O} < 3.2$ Å). Figure 3 shows the time dependence of H-bond number normalized to the number of neighbor molecule pair, which may not satisfy the angular criterion of the hydrogen bond, in CS-I hydrate with different occupancy rate, $I_h$ ice. Each water molecule in the crystal structure of the CS-I hydrate and $I_h$ ice has a tetrahedral ordered environment (short-range order). At high temperatures, the spread in the lengths of hydrogen bonds and the angles of mutual orientation between the neighboring molecules increases due to the thermal motion. However, the number of neighboring molecule pairs, which internal mutual orientation angle is above 30 degrees, was close to 1%. This means that almost all pairs satisfy the H-bond criterion in angle and $N_{H\text{-bonds}}/N_{O-O}$ relation.

The comparison of data presented in Fig. 2 and Fig. 3, allows to conclude that $N_{O-O}$ normalized to $N_{mol}$ before and after melting fluctuates around value of 2. Thus, the main change in the H-bond geometry upon melting is that 10–15% of pairs do not satisfy the angular criterion of the hydrogen bond and the violation of the tetrahedral short-range order takes place. Nevertheless the most of the O···O pairs satisfy the angle H-bond criterion.

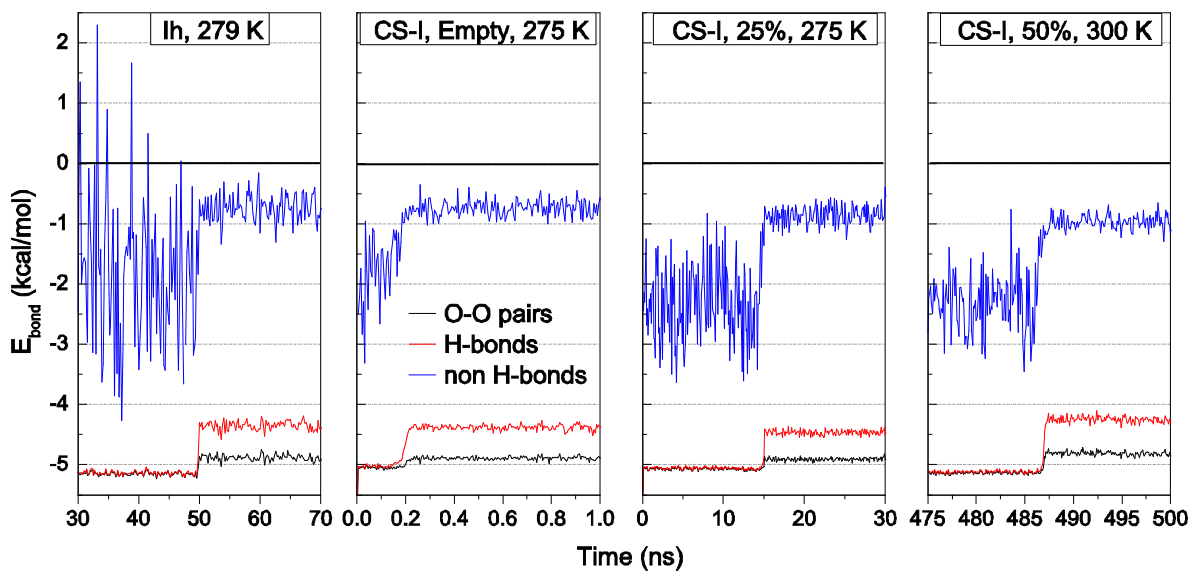

Fig. 4. Time dependence of average interaction energy between O···O pair, H-bonded and non H-bonded pairs in CS-I hydrates with 0%, 25% and 50% occupancy rates at 275 K and 300 K and $I_h$ ice at 279 K.

The sharp change in the geometry of the H-bonds does not lead to a significant degradation in the stability of the H-bond network. Figure 4 shows the time dependence of average interaction energy between every neighbor molecules, molecules that form H-bond and neighbor molecules which satisfy H-bond criterion on distance and do not satisfy the mutual orientation angle H-bond criterion (non H-bond) in CS-I hydrate and $I_h$ ice before and after melting. The obtained results indicates the completeness of the melting process.

The solid phase is characterized by the short-range ordering of each molecule; therefore, before reaching the melting point, the average value of the interaction energy for the H-bonded pairs coincides with the average value for all pairs of neighbor molecules. After melting, the number of non H-bonded pairs significantly increases that leads to a decrease in the average values fluctuation scatter. For the same reason, the average energy value for all pairs increases by ~15%. However, during the melting the binding energy between the H-bonded molecules changes only by 5-7%. This means that the H-bond geometry does not undergo significant changes in the melting process and indicates the preservation of tetrahedral short-range order in the H-bond network after melting.

The insignificant changing in the interaction energy between the H-bonded molecules could be the origin of H-bond network stability due to the fact that the energy of thermal motion in the temperature region of 275 – 300 K is lower than 1 kcal/mol. On average, the interaction energy of any nearby molecules exceeds the value of thermal energy.

The strong fluctuation of the average energy value for pairs of non H-bonded pairs can be explained by thermal motion and ~1% number. Before melting this energy fluctuates close to -2 kcal/mol values, but after melting it fluctuates in the region [-1; -0.8] kcal/mol. In the solid phase, this energy is significantly lower than thermal energy and is responsible for the structure stability in the solid phase. In the liquid phase, this energy is comparable to thermal energy, which explains the ability of its structure to reorganize. The increase of average interaction energy value after melting is due to the further disordering of the short-range order of molecules in the liquid phase.

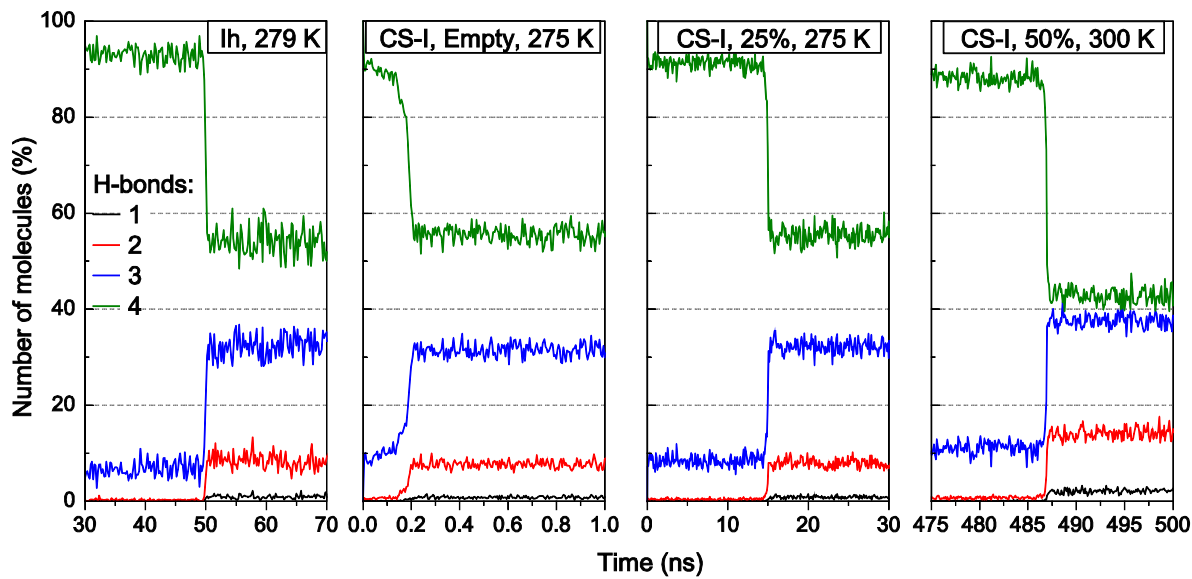

Fig. 5. Dependence of molecule fraction having 1, 2, 3 and 4 H-bonds as a function of time of CS-I hydrates with different occupancy rates at 275 K and 300 K and $I_h$ ice at 279 K.

To determine the short-range order of each molecule in the terms of H-bonding and to study the preservation of tetrahedral type of hydrogen bond network during the solid melting, the time dependencies of the fraction of molecules participating in 1, 2, 3 and 4 H-bonds were obtained and presented in Fig. 5. It is clear that in the solid state the overwhelming majority of molecules has tetrahedral surrounding. Less than 10% of molecules have only 3 water molecules in their short-range order. It can be expected than that thermal motion leads to breaking of the number of H-bonds in solid state, but the presence of $CO_2$ molecules makes hydrate structure stable at higher temperature. In the case of water, only about 50% of the water molecules form 4 hydrogen bonds with their neighbors.

Upon melting the fraction of molecules taking part in the formation of 4 H-bonds *instantly* decreasing by 40-50%, which leads to significant increasing the fraction of molecules that forming 3 H-bonds. This makes tetrahedral H-bonding weaker (with rising of absolute value of full energy) and leads to the local structure changing. The fraction of molecules that taking part in the formation of 3 and 4 H-bonds make up over 85% in total that allows the H-bond network to unite all molecules. The fraction of molecules participating in 2 H-bonds at given temperatures could reach 10%. The number of molecules forming 1 H-bond is insignificantly low. Thus before and after melting H-bond network mostly consists of molecules that form 4 hydrogen bonds – $H_2O$ molecules pairs, which geometry satisfies to H-bond criteria and has a small scatter of lengths and angles. Fraction of molecules forming 4 H-bonds in water decreasing with temperature (and thermal motion) increase.

Figure 2 and Figure 5 show that the presence of small number of $CO_2$ molecules almost does not affect the tetrahedral ordering of H-bond network of solid hydrate. However the presence of cavities leads to instability of ~3-5% of hydrogen bonds at comparing $I_h$ ice and CS-I structures.

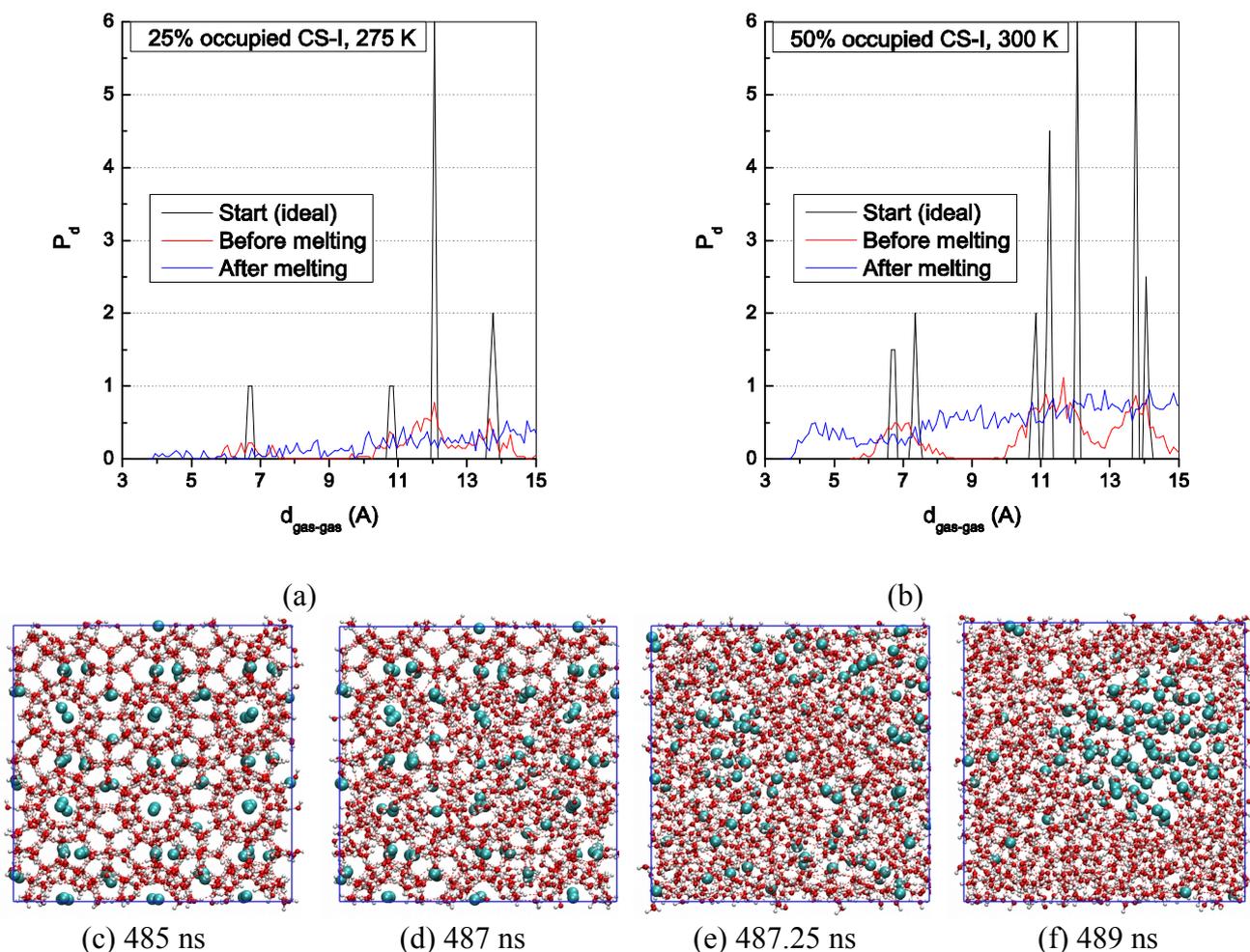

(c) 485 ns  (d) 487 ns  (e) 487.25 ns  (f) 489 ns

Fig. 6. Distribution $P_d$ of $CO_2 \cdots CO_2$ pair lengths $d_{O-O}$ in the CS-I hydrates with 25% (a) and 50% (b) cavity occupancy rate with a 0.01 Å step at the initial (ideal) state, before melting and after melting. Spatial distribution of $CO_2$ molecules before melting (c), at melting point (d) and right after melting point (e) and after 1.75 ns (f) of 50% occupied CS-I hydrate with corresponding time mark.

Figures 6a,b show the distribution of $CO_2 \cdots CO_2$ pair lengths of initial and final configurations and before and after the melting point, respectively. The sharp peaks of the initial state correspond to the regular arrangement of $CO_2$ molecules in hydrate structure. These peaks blur during the heating however $CO_2$ molecules remain inside the deformed cavities. After the melting point the $CO_2$ molecules are distributed evenly throughout the whole volume (Fig. 6e). The small peak in the interval of 4-6 Å as shown in Fig. 6b represent the formation of small $CO_2$ gas bubbles.

## 4. Conclusions

The investigation of the kinetic process of carbon dioxide CS-I hydrate and $I_h$ ice melting at the molecular level showed the instantaneous change of hydrogen bond network as well as short-range order of every $H_2O$ molecule during the solid-liquid transition. This indicates that the nature of dissociation process is a collective phenomenon for ice and carbon dioxide hydrate. The calculated caloric curves clearly showed the jump in the internal energy of the ice, hydrate of $CO_2$ gas with 0%, 25% and 50% cavity occupation ratio in the region 275 K - 300 K. The step by step conducted analysis of sequential structures revealed the instant (less than 1 nanosecond) first-order solid-liquid phase transition at 275, 279 and 300 K.

The analysis showed that every $H_2O$ molecule taking part in the formation of dynamic hydrogen bond network. It was shown that integrity of network was not affected by the phase transition and it persists after the melting point was reached when ~15-17% of H-bonds are broken. The interaction energies of the most of H-bonded pairs are greatly exceed the thermal energy either in solid and liquid phase. The corresponding energies of non H-bonded pairs slightly exceeds and comparable with thermal energies in solid and liquid phase, respectively. Overwhelming majority of O⋯O pairs in solid state are bound by H-bonds defining the solid properties. However, the formation of a hydrogen bond between molecules strongly depends on the distance between the oxygen atoms and the angle of mutual orientation, which can vary greatly with time even in the liquid state.

The presence of cavities in crystalline structure decreases its stability however the increase of $CO_2$ molecules occupied the cavities stabilizes the hydrate structure.

## Acknowledgements

This work of K.G., V.B., R.Z., Y.B., O.S., N.M. and R.G. is supported by the Russian Science Foundation under grant No 18-19-00124 (caloric curves, hydrogen bond structure and energy analysis) that held at the Novosibirsk State University. R.B. and Y.K. are thankful to the Ministry of Education, Culture, Sports, Science, and Technology of Japan (Grant No. 17H03122) for financial support (large-scale molecular dynamics simulation) and are grateful for the continuous support of the crew at the Center for Computer Materials Science at the Institute for Materials Research, Tohoku University, Sendai.

## Author contributions

K.G., Y.B. and V.B. designed the research idea. N. M., O.S. and R. Z. developed software program. R. Z. and K.G. analyzed calculation results. R.B. and Y.K. performed MD simulations on SC in Japan. All authors discussed the results and prepared the manuscript.

## Corresponding authors

Dr. Kirill V. Gets
E-mail address: gets@niic.nsc.ru
Full postal address: Nikolaev Institute of Inorganic Chemistry Siberian Branch of Russian Academy of Sciences, 3, Acad. Lavrentiev Ave., Novosibirsk, 630090, Russia

## Declarations of interest:

none